\documentclass[aps,pra,reprint,superscriptaddress,nofootinbib,amsmath,amssymb]{revtex4-2}

\usepackage{amsthm}
\usepackage[colorlinks=true,allcolors=blue]{hyperref}
\usepackage{graphicx}

\newtheorem{conjecture}{Conjecture}
\newtheorem{proposition}{Proposition}
\newtheorem{lemma}{Lemma}
\newtheorem{corollary}{Corollary}
\theoremstyle{remark}
\newtheorem*{remark}{Remark}

\newcommand{\rr}{\mathbf{r}}
\newcommand{\sss}{\mathbf{s}}
\newcommand{\uu}{\mathbf{u}}
\newcommand{\vv}{\mathbf{v}}
\newcommand{\tr}{\mathrm{Tr}}

\begin{document}

\title{Wigner negativity costs phase-space area: A conjectured refinement of the Lieb--Solovej inequality at R\'enyi order one half}

\author{Christos N. Gagatsos}
\affiliation{School of Electrical, Computing, and Software Engineering, The University of Arizona, Tucson, Arizona 85721, United States of America}

\date{\today}

\begin{abstract}
We conjecture that the Wigner logarithmic negativity of any single-mode state is bounded by the amount by which the R\'enyi-$1/2$ entropy of its Husimi function exceeds the coherent-state value, with equality only for coherent states. If true, this refines the Lieb--Solovej (generalized Wehrl) inequality with a remainder term that is a resource monotone, and it implies that heterodyne statistics alone certify an upper bound on the classical-simulation cost of a state. We show that it suffices to prove the conjecture for pure states, since the negativity is convex and the squared $L^{1/2}$ norm of the Husimi function is concave. We prove the conjecture up to an additive constant $\ln 2$ for all states, and identify where the factor of two is lost: For pure states the Wigner function is the wavefunction of a two-mode state obtained by passive operations, and over generic two-mode wavefunctions our constant is already optimal, so that a proof of the sharp bound must exploit the product structure of that state. Finally, we show with an explicit superposition of the vacuum and a distant coherent state that the analogous inequality fails for the Wehrl entropy and for higher R\'enyi orders, meaning that the conjecture is not a consequence of the majorization relation underlying the Lieb--Solovej theorem.
\end{abstract}

\maketitle

\section{Introduction}
\label{sec:intro}

Phase-space methods describe a quantum state of a bosonic mode by quasi-probability distributions. The Husimi function $Q$ is a genuine probability density, the outcome distribution of heterodyne detection, and its entropies obey sharp lower bounds: Wehrl's conjecture~\cite{Wehrl1979}, proven by Lieb~\cite{Lieb1978} and extended by Lieb and Solovej to all concave functionals~\cite{LiebSolovej2014}, states that no 
state is more concentrated in phase space than a coherent state. 

The Wigner function $W$, on the other hand, can take negative values, and its negativity is a resource that can be quantified by the Wigner logarithmic negativity $\ln\|W\|_1$ which is a monotone in the resource theory of non-Gaussianity~\cite{Albarelli2018,Takagi2018}, and $\|W\|_1^2$ sets the sampling overhead of quasi-probability simulation of quantum circuits~\cite{Mari2012,Pashayan2015}.

Entropic statements have also been proposed for the Wigner function itself. For states with a non-negative Wigner function, the Shannon differential entropy of $W$, known as the Wigner entropy, was conjectured to be bounded from below by its vacuum value $1+\ln\pi$, with equality for pure Gaussian states~\cite{Hertz2017,VanHerstraetenCerf2021}, and a stronger majorization version of this statement was put forward in Ref.~\cite{VanHerstraeten2023majorization}. The Wigner entropy conjecture was proven for several classes of states: passive states~\cite{VanHerstraetenCerf2021}, Wigner non-
negative states supported on the vacuum and the one-photon Fock state~\cite{QianGagatsos2024}, beam-splitter states~\cite{VanHerstraetenCerf2025}, and all Wigner non-negative states whose purity lies below an explicit threshold~\cite{QianGagatsos2026}. For general Wigner-positive states, a lower bound 
with a slightly smaller constant was established, together with the R\'enyi version of the conjecture for R\'enyi orders $\alpha\ge2$~\cite{DiasPrata2023}. Entropic functionals that extend to Wigner functions with negative values have also been 
studied~\cite{CerfHertzVanHerstraeten2024,Pizzimenti2024}. Recently, however, the conjecture was disproved: Wigner-positive states whose Wigner entropy lies strictly below the vacuum value were constructed in Refs.~\cite{He2026,VanHerstraetenCerfChabaud2026}, and the second of these also disproves the majorization conjecture. 

In contrast, the conjecture we propose here concerns the Husimi function, for which the corresponding entropic lower bound is a theorem~\cite{Lieb1978,LiebSolovej2014}, and it relates the excess of that entropy to the Wigner negativity.

The Lieb--Solovej bound says that a state cannot be more concentrated than a coherent state but it says nothing about how much more spread out a state must be in order to be non-classical. In this paper we propose that the price, measured by the R\'enyi entropy of order $1/2$ of the Husimi function, is exactly the negativity:
\begin{equation}
\ln\|W\|_1 \le h_{1/2}(Q)-\ln(8\pi),
\label{eq:conjintro}
\end{equation}
for every single-mode state, with equality only for coherent states (precise definitions are given in Sec.~\ref{sec:setting}). We see four reasons for interest in this statement. First, it would be a refined Wehrl-type inequality, $h_{1/2}(Q)\ge\ln(8\pi)+\ln\|W\|_1$, whose deficit is controlled by an operational resource rather than by a distance to the set of coherent states. Second, $e^{h_{1/2}(Q)}$ is an effective phase-space area, equal to $8\pi$ for a coherent state, so that Eq.~\eqref{eq:conjintro} reads ``negativity costs area'': $\|W\|_1$ is at most the area occupied by the state in units of the coherent-state area. Third, operationally, the statistics of a single Gaussian measurement would certify an upper bound on the classical-simulation overhead $\|W\|_1^2$, and on any conversion rate limited by the logarithmic negativity, without tomography. Fourth, Lieb's inequalities bound $\|W\|_p$ from below for $p<2$, and Eq.~\eqref{eq:conjintro} would be a complementary upper bound at $p=1$.

This paper is organized as follows. In Sec.~\ref{sec:setting} we set the notation and state the conjecture. In Sec.~\ref{sec:pure} we show that it suffices to prove it for pure states. Exploiting the results of the previous section, we test numerically the conjecture for random pure states. In Sec.~\ref{sec:weak} we prove the conjecture up to an additive constant $\ln2$ and explain where the factor of two is lost, including a structural reason why the sharp constant cannot be obtained by arguments that treat the Wigner function as a generic function. In Sec.~\ref{sec:majorization} we show that the conjecture is not a majorization statement, and in Sec.~\ref{sec:conclusions} we summarize, and discuss further the conjecture and related open questions.

\section{Setting and the conjecture}
\label{sec:setting}

We take $\hbar=1$, $[\hat x,\hat p]=i$, and write $\rr=(x,p)$ for a phase-space point, with $\alpha=(x+ip)/\sqrt2$ the corresponding coherent amplitude and $|\rr\rangle\equiv|\alpha\rangle$ the coherent state centred at $\rr$. For a single-mode state $\rho$ the Wigner and Husimi functions are
\begin{eqnarray}
W(\rr)&=&\frac1\pi \tr \big[\rho \hat D(\rr) \hat\Pi \hat D^\dagger(\rr)\big],\\
Q(\rr)&=&\frac{1}{2\pi} \langle\rr|\rho|\rr\rangle ,
\label{eq:defs}
\end{eqnarray}
where $\hat D(\rr)$ is the displacement operator and $\hat\Pi=(-1)^{\hat n}$ is the photon-number parity. Both quasi-probabilities are normalized, $\int d^2\rr W=\int d^2\rr Q=1$ with $d^2\rr=dx dp$.

Two functionals are of interest. The first is the $L^1$ norm of the Wigner function,
\begin{equation}
\|W\|_1=\int d^2\rr |W(\rr)| \ge 1,
\end{equation}
which equals $1$ exactly when $W\ge0$ and its logarithm is the Wigner logarithmic negativity. The second is the R\'enyi entropy of order $1/2$ of the Husimi function,
\begin{equation}
h_{1/2}(Q)=2\ln\!\int d^2\rr \sqrt{Q(\rr)} .
\end{equation}
The Lieb--Solovej theorem~\cite{Lieb1978,LiebSolovej2014} gives
\begin{equation}
h_{1/2}(Q) \ge \ln(8\pi),
\label{eq:LS}
\end{equation}
with equality for coherent states.

\begin{conjecture}
For every single-mode state $\rho$,
\begin{equation}
\begin{split}
\int d^2\rr |W(\rr)|& \le \frac{1}{8\pi}\left(\int d^2\rr \sqrt{Q(\rr)}\right)^{2},\\
\textup{or}\qquad\ln\|W\|_1& \le h_{1/2}(Q)-\ln(8\pi),
\end{split}
\label{eq:conj}
\end{equation}
with equality if and only if $\rho$ is a coherent state.
\end{conjecture}

Read from right to left, Eq.~\eqref{eq:conj} is the Lieb--Solovej bound~\eqref{eq:LS} with a non-negative remainder $\ln\|W\|_1$ that vanishes on Wigner-positive states.~Therefore, for Wigner-positive states the conjecture becomes the Lieb--Solovej theorem.

\begin{remark}
The two sides of Eq.~\eqref{eq:conj} have different symmetries. Under a Gaussian unitary $\hat U$, acting on the quadratures as $\hat U^\dagger(\hat x,\hat p)^T\hat U=S(\hat x,\hat p)^T+\mathbf d$ with $S$ symplectic, the Wigner function transforms covariantly, $W_{\hat U\rho\hat U^\dagger}(\rr)=W_\rho(S^{-1}(\rr-\mathbf d))$, so that $\|W\|_1$ is invariant. The right-hand side is invariant under displacements and rotations, but not under squeezing. For instance, a squeezed vacuum has $\|W\|_1=1$ but $h_{1/2}(Q)>\ln(8\pi)$. A symplectically invariant bound is obtained by replacing the right-hand side by its infimum over $\hat U\rho\hat U^\dagger$, i.e.\ by replacing heterodyne detection with the best general-dyne measurement. In that form all pure Gaussian states saturate the bound. This version is not stronger than Eq.~\eqref{eq:conj}: applying Eq.~\eqref{eq:conj} to $\hat U\rho\hat U^\dagger$ and using the invariance of $\|W\|_1$ gives $\ln\|W_\rho\|_1\le h_{1/2}(Q_{\hat U\rho\hat U^\dagger})-\ln(8\pi)$ for every $\hat U$, and the infimum can then be taken on the right. The two forms are therefore equivalent, and only Eq.~\eqref{eq:conj} needs to be proven.
\end{remark}

\section{Pure states suffice}
\label{sec:pure}

\begin{proposition}
\label{prop:pure}
If Eq.~\eqref{eq:conj} holds for all pure states, it holds for all states. Moreover, if among pure states only coherent states attain equality in \eqref{eq:conj}, then every mixed state satisfies \eqref{eq:conj} with strict inequality.
\end{proposition}

\begin{proof}
Write $M(\rho)=\int d^2\rr \sqrt{Q_\rho(\rr)}$, so that the right-hand side of the first form of Eq.~\eqref{eq:conj} is $M(\rho)^2/8\pi$. Let $\rho=\sum_kp_k\rho_k$ be a decomposition into pure states $\rho_k$ with $p_k>0$ and $\sum_k p_k=1$. Since $W$ is linear in the state, the triangle inequality gives
\begin{equation}
\|W_\rho\|_1\le\sum_kp_k \|W_{\rho_k}\|_1 .
\label{eq:convex}
\end{equation}
For the right-hand side we show that $\rho\mapsto M(\rho)^2$ is concave. Since $Q_\rho=\sum_kp_kQ_{\rho_k}$ with $p_k>0$, every $Q_{\rho_k}$ vanishes wherever $Q_\rho$ does, and all integrals below can be restricted to the set where $Q_\rho>0$. By the Cauchy--Schwarz inequality,
\begin{equation}
\begin{split}
M(\rho_k)^2&=\Big(\int d^2\rr \frac{\sqrt{Q_{\rho_k}}}{Q_\rho^{1/4}} Q_\rho^{1/4}\Big)^2\\
&\le\int d^2\rr \frac{Q_{\rho_k}}{\sqrt{Q_\rho}}\int d^2\rr \sqrt{Q_\rho}\\
&=M(\rho)\int d^2\rr \frac{Q_{\rho_k}}{\sqrt{Q_\rho}} .
\end{split}
\label{eq:CSmix}
\end{equation}
Multiplying by $p_k$ and summing over $k$, $\sum_kp_kQ_{\rho_k}=Q_\rho$ gives
\begin{equation}
\sum_kp_k M(\rho_k)^2\le M(\rho)\int d^2\rr \sqrt{Q_\rho}=M(\rho)^2 .
\label{eq:concave}
\end{equation}
If Eq.~\eqref{eq:conj} holds for every $\rho_k$, then Eqs.~\eqref{eq:convex} and~\eqref{eq:concave} give
\begin{equation}
\begin{split}
\|W_\rho\|_1&\le\sum_kp_k\|W_{\rho_k}\|_1\\
&\le\frac{1}{8\pi}\sum_kp_kM(\rho_k)^2\le\frac{1}{8\pi}M(\rho)^2 .
\end{split}
\end{equation}
For the equality case, equality in Eq.~\eqref{eq:concave} requires equality in Eq.~\eqref{eq:CSmix} for every $k$, i.e., $\sqrt{Q_{\rho_k}}\propto\sqrt{Q_\rho}$ and by normalization $Q_{\rho_k}=Q_\rho$ for all $k$. Since the Husimi function determines the state, all $\rho_k$ coincide with $\rho$, which is then pure.
\end{proof}

The proof also shows that $\|W_\rho\|_1-M(\rho)^2/8\pi$ is a convex function of $\rho$ (as the difference of a convex and a concave function) Therefore, its maximum over any convex set of states is attained at extreme points, which is why violations must be sought among pure states.

\section{Numerical Evaluations}\label{sec:numerics}
\begin{figure*}[t!]
\centering
\begin{minipage}[b]{0.48\textwidth}
  \centering
  \includegraphics[width=\linewidth]{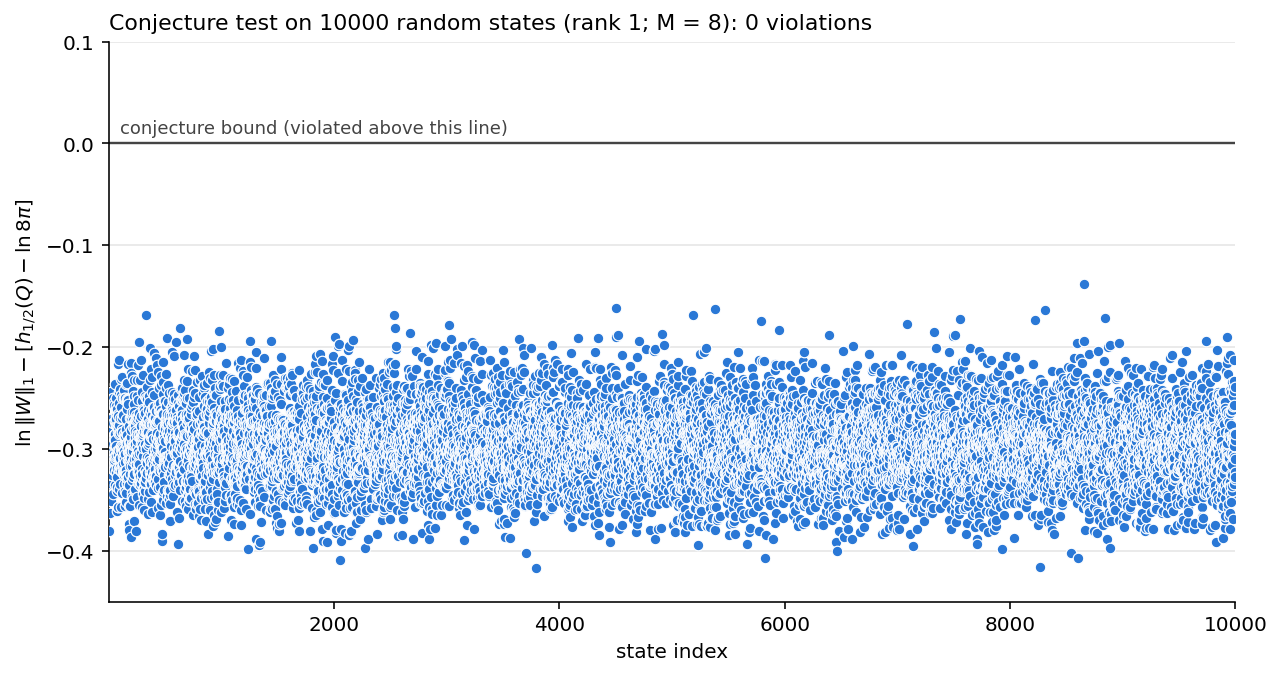}\\[-2pt]
  {\small (a)}
\end{minipage}\hfill
\begin{minipage}[b]{0.48\textwidth}
  \centering
  \includegraphics[width=\linewidth]{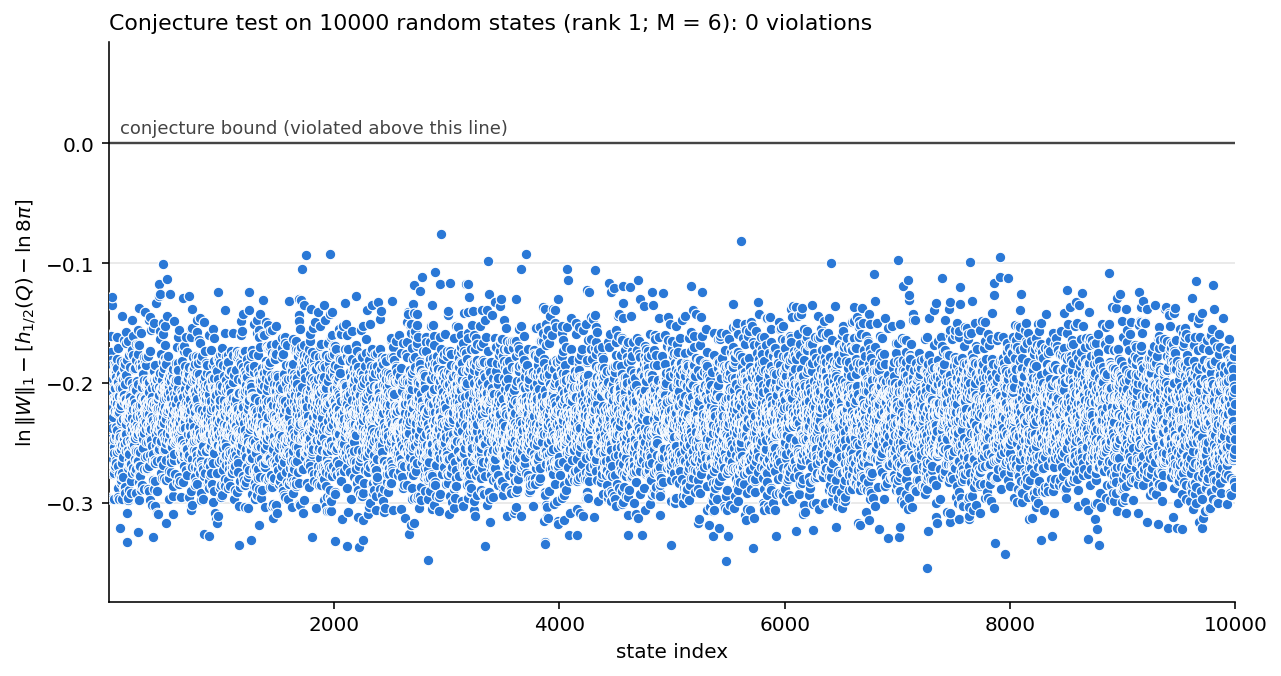}\\[-2pt]
  {\small (b)}
\end{minipage}

\vspace{8pt}

\begin{minipage}[b]{0.48\textwidth}
  \centering
  \includegraphics[width=\linewidth]{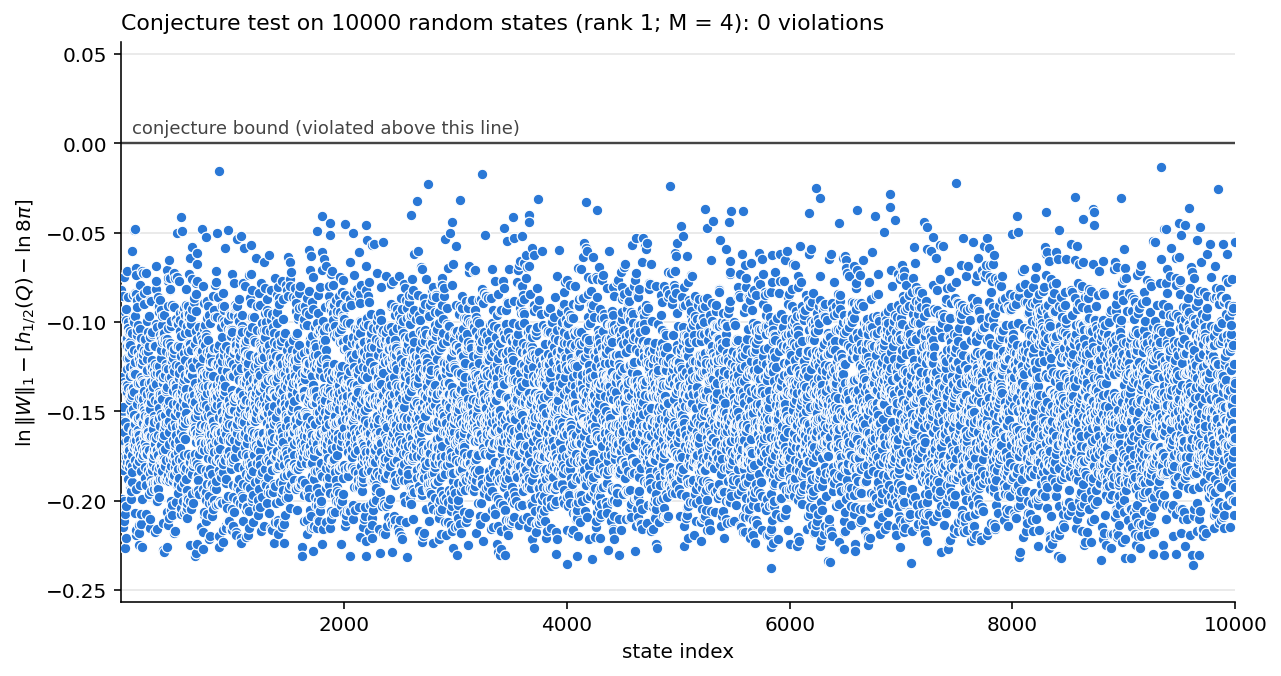}\\[-2pt]
  {\small (c)}
\end{minipage}\hfill
\begin{minipage}[b]{0.48\textwidth}
  \centering
  \includegraphics[width=\linewidth]{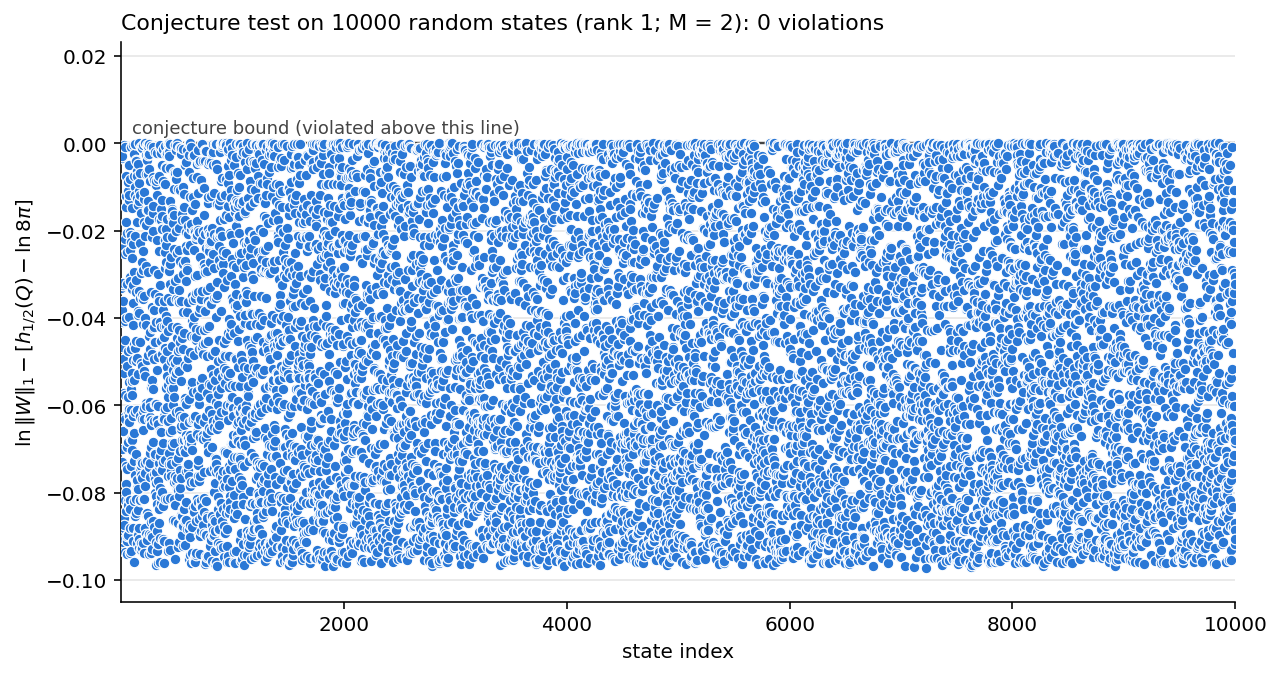}\\[-2pt]
  {\small (d)}
\end{minipage}
\caption{The difference $\ln\|W\|_1 - [h_{1/2}(Q)-\ln(8\pi)]$ of $10000$ random Fock states' superpositions on $\{|0\rangle,\ldots,|M-1\rangle\}$ for (a) $M=8$, (b) $M=6$, (c) $M=4$ and (d) $M=2$. The $x-$axis labels the random state. Any point above the horizontal line set on $0$ would indicate a violation of the conjecture.}
\label{fig:justRandom}
\end{figure*}

\begin{figure*}[t!]
\centering
\begin{minipage}[b]{0.48\textwidth}
  \centering
  \includegraphics[width=\linewidth]{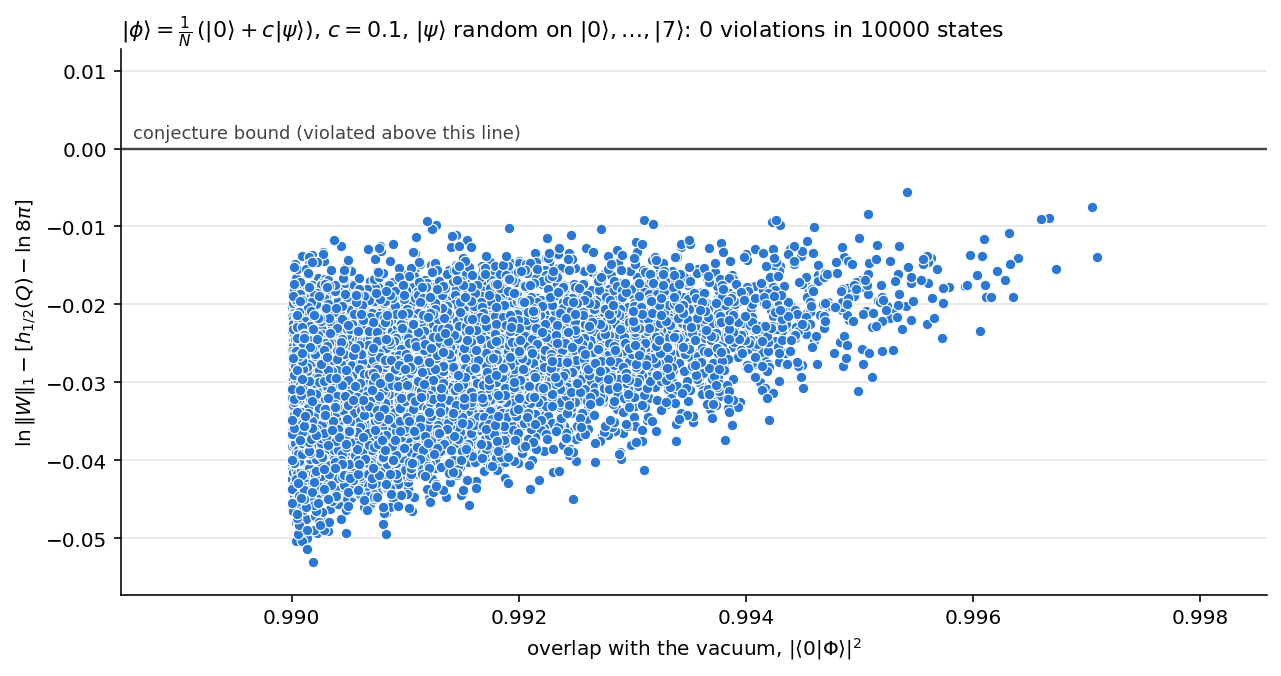}\\[-2pt]
  {\small (a)}
\end{minipage}\hfill
\begin{minipage}[b]{0.48\textwidth}
  \centering
  \includegraphics[width=\linewidth]{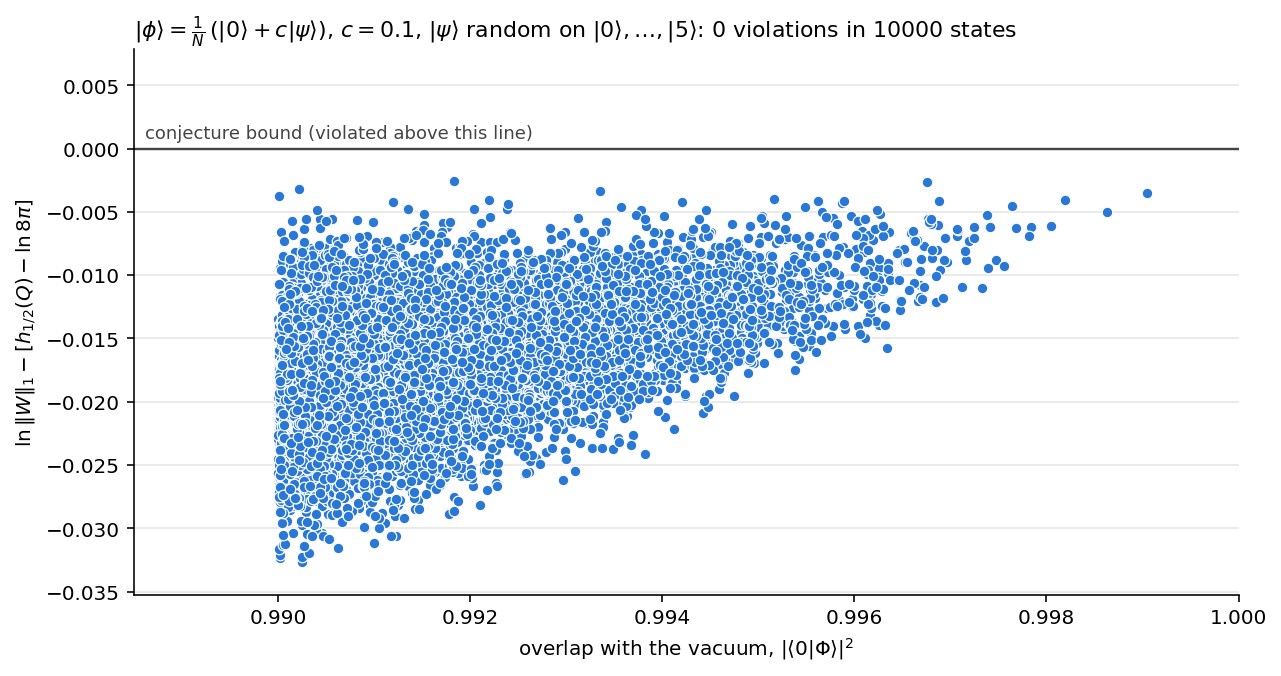}\\[-2pt]
  {\small (b)}
\end{minipage}

\vspace{8pt}

\begin{minipage}[b]{0.48\textwidth}
  \centering
  \includegraphics[width=\linewidth]{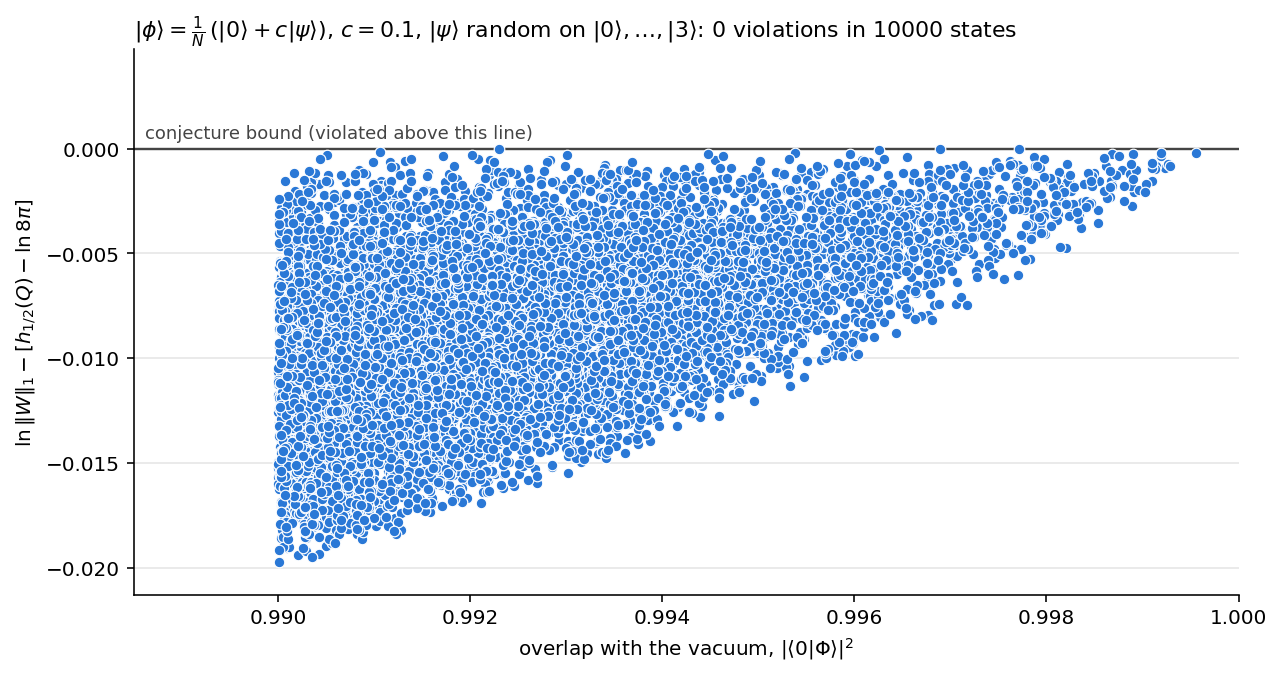}\\[-2pt]
  {\small (c)}
\end{minipage}\hfill
\begin{minipage}[b]{0.48\textwidth}
  \centering
  \includegraphics[width=\linewidth]{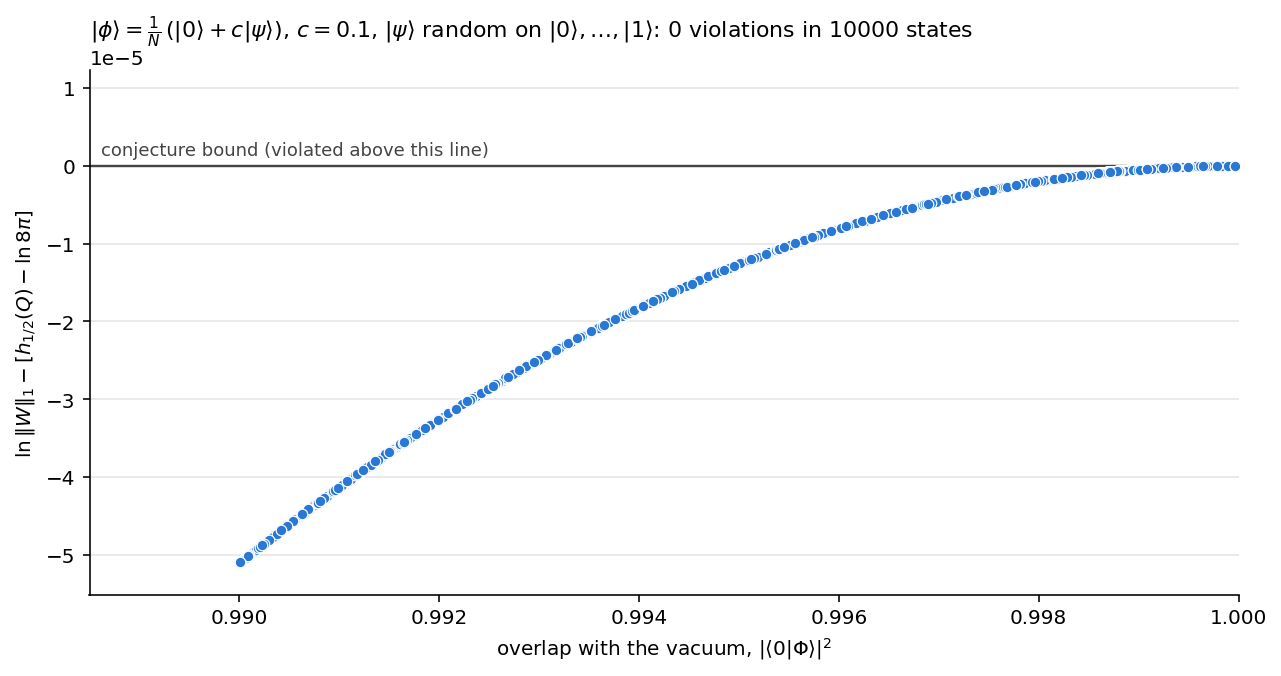}\\[-2pt]
  {\small (d)}
\end{minipage}
\caption{The difference $\ln\|W\|_1 - [h_{1/2}(Q)-\ln(8\pi)]$ of $10000$ states of the form $|\phi\rangle=N^{-1} (|0\rangle+|\psi\rangle)$, where $|\psi\rangle$ are random Fock states' superpositions on $\{|0\rangle,\ldots,|M-1\rangle\}$ for (a) $M=8$, (b) $M=6$, (c) $M=4$ and (d) $M=2$. Said difference is plotted as a function of the probability $|\langle 0|\phi\rangle|^2$. Any point above the horizontal line set on $0$ would indicate a violation of the conjecture. The smallest difference observed is $-0.00005$.}
\label{fig:psiRandom}
\end{figure*}

Since we have established that it is enough to prove the conjecture for pure states, in this section we run numerical tests for two kinds of states: (i) Random generated pure states and (ii) states of the form $|\phi\rangle=N^{-1}(|0\rangle)+|\psi\rangle)$, where $|\psi\rangle$ is random. The motivation of the latter kind is that the random states of kind (i), show that the conjecture is truly tested, i.e.,  the quantity $\ln\|W\|_1 - [h_{1/2}(Q)-\ln(8\pi)]$ (see Eq. \eqref{eq:conj}) tends to zero from below, for states with small Fock expansion length and for high probability $|\langle 0|\phi\rangle|^2$.

The states of kind (i) and kind (ii) are drawn uniformly (Haar-random) from the pure states on the truncated Fock space spanned by $|0\rangle,\ldots,|M\rangle$ (we allow the $|0\rangle$ in $|\psi\rangle$). To do this, the $M$ amplitudes are drawn as independent standard complex Gaussian random variables (real and imaginary parts independent with zero mean and unit variance), and the resulting vector is then normalized.

We show the numerical results in Figs. \ref{fig:justRandom} and \ref{fig:psiRandom}. Several thousands of states of both kinds have been tested and we did not find any violations. Here we present two kinds of plots. For states of kind (i) we plot the the difference $\ln\|W\|_1 - [h_{1/2}(Q)-\ln(8\pi)]$ per random state, while for the states of kind (ii) we plot the same difference as a function of the probability $|\langle 0|\phi\rangle|^2$.

\section{A weaker version of the conjecture}
\label{sec:weak}

\begin{proposition}
\label{prop:weak}
For every single-mode state $\rho$,
\begin{equation}
\begin{split}
\int d^2\rr |W(\rr)|& \le \frac{1}{4\pi}\left(\int d^2\rr \sqrt{Q(\rr)}\right)^{2},\\
\textup{or}\qquad\ln\|W\|_1& \le h_{1/2}(Q)-\ln(4\pi).
\end{split}
\label{eq:weak}
\end{equation}
\end{proposition}

\begin{proof}
Evaluate the trace in Eq.~\eqref{eq:defs} with the coherent-state resolution of the identity, $\int d^2\sss |\sss\rangle\langle\sss|/2\pi=\hat 1$. Since $\hat\Pi|\sss\rangle=|-\sss\rangle$ and displacements map coherent states to coherent states up to a phase,
\begin{equation}
W(\rr)=\frac1\pi\int\frac{d^2\sss}{2\pi} e^{i\varphi(\rr,\sss)} \langle\rr+\sss|\rho|\rr-\sss\rangle ,
\label{eq:parity}
\end{equation}
with $\varphi$ real and explicitly $\varphi(\rr,\sss)=xs_p-ps_x$ (Appendix~\ref{app:phase}). Taking the modulus and moving it inside the integral (triangle inequality) removes the phase. Positivity of $\rho$ and the Cauchy--Schwarz inequality then give $|\langle\uu|\rho|\vv\rangle|\le\sqrt{\langle\uu|\rho|\uu\rangle\langle\vv|\rho|\vv\rangle}=2\pi\sqrt{Q(\uu)Q(\vv)}$, hence
\begin{equation}
|W(\rr)| \le \frac1\pi\int d^2\sss \sqrt{Q(\rr+\sss) Q(\rr-\sss)} .
\label{eq:pointwise}
\end{equation}
Integrating over $\rr$ and changing variables to $\uu=\rr+\sss$, $\vv=\rr-\sss$, for which $d^2\rr d^2\sss=\tfrac14 d^2\uu d^2\vv$, the double integral factorizes and Eq.~\eqref{eq:weak} follows.
\end{proof}

The proof uses nothing beyond positivity of $\rho$, so it holds for mixed states directly. Combined with Eq.~\eqref{eq:LS}, we establish the following,
\begin{equation}
h_{1/2}(Q) \ge \max\big\{\ln(8\pi), \ln(4\pi)+\ln\|W\|_1\big\}.
\label{eq:max}
\end{equation}
The second entry improves on the Lieb--Solovej bound only when $\|W\|_1>2$ (e.g. for high Fock states). Proposition~\ref{prop:weak} is thus a complementary bound for strongly negative states, and not yet a refinement of the Lieb--Solovej inequality for all states. The missing $\ln2$ is what the conjecture would add.

\subsection{How the factor of two is lost}

Two inequalities were used in the proof: the triangle inequality, which discards the phase in Eq.~\eqref{eq:parity}, and the Cauchy--Schwarz inequality for the matrix element. For pure states the second is an equality, $|\langle\uu|\psi\rangle\langle\psi|\vv\rangle|=|\langle\uu|\psi\rangle| |\langle\psi|\vv\rangle|$, so for pure states (to which the conjecture reduces by Proposition~\ref{prop:pure}) the entire loss comes from discarding the phase. The matrix element in Eq.~\eqref{eq:parity} is positive, and for vacuum the right-hand side of Eq.~\eqref{eq:pointwise} equals $e^{-(x^2+p^2)/2}/\pi$, in place of the true $W=e^{-(x^2+p^2)}/\pi$. The bound is tight at the origin and nowhere else: For $\rr\neq0$ the phase $e^{i(xs_p-ps_x)}$ oscillates in $\sss$, and the resulting cancellation is what makes the Wigner function decay twice as fast. Integrated over phase space, this is exactly the factor of two between Eqs.~\eqref{eq:weak} and~\eqref{eq:conj}. A proof of the conjecture must retain this interference.

There is also a structural reason why the phase cannot be recovered by a generic argument. Let us see this in four steps, using that it suffices to consider a pure state $|\psi\rangle$.
\begin{enumerate}
\item \emph{The Wigner function is a wavefunction.} Let $|\bar\psi\rangle$ have the complex-conjugate position wavefunction of $|\psi\rangle$, and let $|\Xi\rangle$ be the two-mode state obtained from $|\bar\psi\rangle\otimes|\psi\rangle$ by a balanced beam splitter followed by a Fourier transform on the second mode. Then $W(x,p)=\pi^{-1/2} \Xi(\sqrt2x,\sqrt2p)$, where $\Xi(x,y)$ is the position wavefunction of $|\Xi\rangle$ (Appendix~\ref{app:twomode}). In particular $\|W\|_1$ is, up to a constant, the position-space $L^1$ norm of $\Xi$.
\item \emph{The right-hand side is a coherent-state norm of the same wavefunction.} For any $n$-mode state let $N(\Phi)=\int d^{2n}\alpha \pi^{-n}|\langle\boldsymbol\alpha|\Phi\rangle|$ be the $L^1$ norm of its coherent-state amplitude. For one mode, $N(\psi)=(2\pi)^{-1/2}\int d^2\rr\sqrt Q$. The beam splitter and the Fourier transform are passive, i.e., they map coherent states to coherent states, so they only rotate the argument of the coherent amplitude, and $N$ is unchanged. Hence $N(\Xi)=N(\bar\psi\otimes\psi)=N(\psi)^2=(\int\!\sqrt Q)^2/2\pi$ (Appendix~\ref{app:passive}).
\item \emph{Both bounds become a comparison to the same quantity.} Inserting these identities, the conjecture and Proposition~\ref{prop:weak} read, for pure states,
\begin{equation}
\begin{aligned}
\|\Xi\|_{L^1(\mathbb R^2)}&\le\tfrac{\sqrt\pi}{2} N(\Xi)&&\text{(conjecture)},\\
\|\Xi\|_{L^1(\mathbb R^2)}&\le\sqrt\pi N(\Xi)&&\text{(Proposition~\ref{prop:weak})},
\end{aligned}
\end{equation}
i.e.\ both compare the position-space $L^1$ norm of a two-mode wavefunction with the $L^1$ norm of its coherent-state amplitude. The two-mode vacuum saturates the first.
\item \emph{For generic wavefunctions the weak constant is optimal.} The norm $N$ treats position and momentum on an equal footing, whereas the position-space $L^1$ norm depends on position only. Therefore, a wavefunction that is much broader than a coherent state in position, and hence narrower in momentum, has a large position-space $L^1$ norm relative to $N$. For the product of two squeezed vacua of position width $s$ the ratio $\|\Xi\|_{L^1}/N(\Xi)$ equals $\sqrt\pi s^2/(1+s^2)$, i.e., it exceeds the conjectured $\sqrt\pi/2$ for every $s>1$ and tends to $\sqrt\pi$ as $s\to\infty$. Conversely, $\|\Phi\|_{L^1}\le\sqrt\pi N(\Phi)$ holds for every two-mode wavefunction (Appendix~\ref{app:sharp}).
\end{enumerate}
Consequently, the constant $1/4\pi$ of Proposition~\ref{prop:weak} is the best that any argument can give if it treats $\Xi$ as an arbitrary two-mode wavefunction, and a proof of the conjecture must use that $|\Xi\rangle$ comes from the product $|\bar\psi\rangle\otimes|\psi\rangle$. This structure is what forbids the bad cases above: a Wigner function cannot be broad in both $x$ and $p$, and indeed the wavefunctions that approach the weak constant would correspond to $W\propto e^{-(x^2+p^2)/s^2}$, which for $s\ne1$ is not the Wigner function of a pure state (Appendix~\ref{app:sharp}).

\section{The conjecture is not a majorization statement}
\label{sec:majorization}

The Lieb--Solovej theorem is not specific to the order $1/2$, i.e., it states that the Husimi function of the vacuum majorizes that of every state, so that for every R\'enyi order $\alpha$ the entropy excess
\begin{equation}
\begin{split}
\Delta_\alpha(\rho)&=h_\alpha(Q)-h_\alpha(Q_{\rm vac}) \ge 0,\\
h_\alpha(Q)&=\frac{1}{1-\alpha}\ln\!\int d^2\rr Q^\alpha ,
\end{split}
\label{eq:LSalpha}
\end{equation}
is non-negative. For $\alpha=1$ is the Wehrl entropy, $\alpha=\infty$ gives $h_\infty(Q)=-\ln\max Q$. The right-hand side of Eq.~\eqref{eq:conj} is $\Delta_{1/2}$. A natural way to read the conjecture, ``negativity costs area'', would be a majorization statement: $Q$ is majorized by the dilated vacuum $g_\lambda(\rr)=e^{-|\rr|^2/2\lambda}/2\pi\lambda$ with $\lambda=\|W\|_1$. Since $h_\alpha(g_\lambda)-h_\alpha(Q_{\rm vac})=\ln\lambda$ for every $\alpha$, this would give $\ln\|W\|_1\le\Delta_\alpha(\rho)$ for all orders at once. In particular, at $\alpha=\infty$ it would give $\max Q\le\max g_\lambda=1/2\pi\lambda$, i.e.
\begin{equation}
\|W\|_1 \le \frac{1}{F_{\max}(\rho)},\qquad F_{\max}(\rho)=\max_{\rr} \langle\rr|\rho|\rr\rangle ,
\label{eq:Finf}
\end{equation}
a bound of the negativity by the inverse of the best coherent-state fidelity.

However, this is false. Consider the unbalanced superposition of the vacuum and a distant coherent state,
\begin{equation}
|\psi\rangle\propto|0\rangle+t |\beta\rangle,\qquad t=0.2,\quad\beta=3\sqrt2 ,
\label{eq:catcounter}
\end{equation}
i.e.\ $|\beta\rangle$ is centred at $\rr=(6,0)$. Since $\langle0|\beta\rangle=e^{-9}$ is negligible, $F_{\max}=1/(1+t^2)$ up to corrections of order $e^{-9}$, so the right-hand side of Eq.~\eqref{eq:Finf} is $1.04$, whereas a direct evaluation gives $\|W\|_1=1.227$. The same state also violates the corresponding inequality at the other orders above $1/2$ that we evaluated, while it satisfies the conjecture. With $\ln\|W\|_1=0.204$, the entropy excesses are
\begin{center}
\begin{tabular}{lcccc}
$\alpha$ & $1/2$ & $1$ & $2$ & $\infty$\\ \hline
$\Delta_\alpha(\psi)$ & $0.304$ & $0.161$ & $0.077$ & $0.039$\\
\end{tabular}
\end{center}
The mechanism is the following. The interference fringes between the two components carry a negativity of order $t$, while the weak far component changes the concentration of $Q$ only at order $t^2$ for $\alpha\ge1$, and at order $t^{2\alpha}$ in general. Only at $\alpha=1/2$, where $\int\!\sqrt Q$ is the $L^1$ norm of the coherent amplitude $\langle\rr|\psi\rangle$ and hence linear in it, does the right-hand side respond at order $t$, matching the bilinear dependence of $W$ on $\psi$.

Hence the conjecture does not extend to the Wehrl entropy, and it cannot be obtained from the majorization of $Q$ by a dilated vacuum, or by any reference density whose R\'enyi excesses at all orders are at least $\ln\|W\|_1$. A proof must use the relation between $W$ and the coherent amplitude itself, and not only the distribution of the values of $Q$.

\section{Conclusions}
\label{sec:conclusions}

We have proposed that the Wigner logarithmic negativity of a single-mode state is bounded by the excess of the R\'enyi-$1/2$ entropy of its Husimi function over the coherent-state value, Eq.~\eqref{eq:conj}. We showed that it suffices to prove the bound for pure states, proved it for all states up to an additive constant $\ln2$, and traced the missing constant to the phase discarded by the triangle inequality. The reformulation of the Wigner function as the wavefunction of a two-mode state shows that the constant we obtain is already optimal over generic wavefunctions, so the sharp bound, if true, rests on the product structure $|\bar\psi\rangle\otimes|\psi\rangle$. Finally, the conjecture is specific to the order $1/2$ and is not a consequence of majorization.

The last point deserves an informal comment, since majorization is the mechanism behind the Lieb--Solovej theorem and its modern proofs. Majorization constrains only the distribution of the values of $Q$, i.e.\ how much phase-space area the state occupies at each level of concentration. The Wigner function, in contrast, depends on the coherent amplitude itself, including its phase, through the interference term of Eq.~\eqref{eq:parity}. This is why the order $1/2$ is singled out: $\int\!\sqrt Q$ is the $L^1$ norm of the amplitude, which responds linearly to a weak superposed component exactly as the negativity does. Heuristically, the same dilute superpositions as in Sec.~\ref{sec:majorization}, with the separation growing as $t\to0$, suggest that the analogous inequality fails for every order $\alpha>1/2$, since the negativity is of order $t$ and $\Delta_\alpha$ of order $t^{2\alpha}$. The case of large Fock states suggest that it also fails for sufficiently small orders, roughly $\alpha\lesssim0.17$, because their Wigner negativity grows as $1/2\ln n$ while $\Delta_\alpha$ falls below it by an $\alpha$-dependent constant. If so, any inequality of this type can hold only in a window of R\'enyi orders whose upper end is $1/2$.

Apart from proving or disproving the conjecture itself, further questions arise: (i) Whether coherent states are at least local maximizers of $\|W\|_1/(\int\!\sqrt Q )^2$. (ii) Whether the sharp constant can be reached by an inequality that retains part of the phase, for instance by a stationary-phase estimate of Eq.~\eqref{eq:parity}, in which the phase of the coherent amplitude is related to $\ln Q$ through the Cauchy--Riemann equations of the Bargmann representation. (iii) Whether the multimode version holds with the constant $(8\pi)^{-n}$ and the weak form holds with $(4\pi)^{-n}$, and Proposition~\ref{prop:pure} reduces the problem to pure states for any number of modes. (iv) Whether the bound holds at R\'enyi orders between the lower threshold mentioned above and $1/2$ as this does not follow from the case $\alpha=1/2$.
\newline
\section*{Acknowledgments}
The author acknowledges useful discussions with Nicolas Cerf (ULB, Belgium) and Zacharie Van Herstraeten (INRIA, France). The author acknowledges the use of Anthropic's Claude (model Opus 5.5) in building and debugging the Python code used in Section \ref{sec:numerics}. What the code should do, how to do it, and how to present the data was decided by the author. The author takes full responsibility for the numerical results.

\onecolumngrid
\appendix

\section{The phase in the parity representation and the vacuum example}
\label{app:phase}
This Appendix collects known results from the literature and is given here for completeness.

We use $\hat D(\rr)=\exp[i(p\hat x-x\hat p)]=\exp(\alpha\hat a^\dagger-\bar\alpha\hat a)$ with $\alpha=(x+ip)/\sqrt2$, and $|\rr\rangle=\hat D(\rr)|0\rangle$. From $e^{A+B}=e^Ae^Be^{-[A,B]/2}$ one has $\hat D(\rr)=e^{-ipx/2}e^{ip\hat x}e^{-ix\hat p}$, so that in the position representation
\begin{equation}
\big(\hat D(\rr)\psi\big)(y)=e^{ipy-ipx/2} \psi(y-x),
\qquad
\langle y|\rr\rangle=\pi^{-1/4}\exp\!\Big[-1/2(y-x)^2+ipy-\tfrac{i}{2}px\Big].
\label{eq:cohwf}
\end{equation}
Displacements compose as $\hat D(\rr)\hat D(\sss)=e^{\frac{i}{2}(p s_x-x s_p)}\hat D(\rr+\sss)$, where $\sss=(s_x,s_p)$. Hence
\begin{equation}
\hat D(\rr)|\sss\rangle=e^{\frac{i}{2}(p s_x-x s_p)} |\rr+\sss\rangle .
\label{eq:Dcoh}
\end{equation}

\begin{lemma}[Explicit form of Eq.~\eqref{eq:parity}]
\label{lem:phase}
For every state $\rho$,
\begin{equation}
W(\rr)=\frac1\pi\int\frac{d^2\sss}{2\pi} e^{i(x s_p-p s_x)} \langle\rr+\sss|\rho|\rr-\sss\rangle ,
\label{eq:parityexplicit}
\end{equation}
i.e.\ Eq.~\eqref{eq:parity} holds with $\varphi(\rr,\sss)=x s_p-p s_x$.
\end{lemma}

\begin{proof}
By cyclicity of the trace and the resolution of the identity, $\tr[\rho\hat D\hat\Pi\hat D^\dagger]=\tr[\hat D^\dagger\rho\hat D\hat\Pi]=\int\frac{d^2\sss}{2\pi}\langle\sss|\hat D^\dagger(\rr) \rho \hat D(\rr) \hat\Pi|\sss\rangle$. Since $\hat\Pi\hat a\hat\Pi=-\hat a$ and $\hat\Pi|0\rangle=|0\rangle$, we have $\hat\Pi|\sss\rangle=|-\sss\rangle$. Equation~\eqref{eq:Dcoh} gives $\hat D(\rr)|-\sss\rangle=e^{-\frac{i}{2}(p s_x-x s_p)}|\rr-\sss\rangle$ and $\langle\sss|\hat D^\dagger(\rr)=e^{-\frac{i}{2}(p s_x-x s_p)}\langle\rr+\sss|$. Multiplying the two phases gives $e^{-i(p s_x-x s_p)}$, which is Eq.~\eqref{eq:parityexplicit} after inserting the prefactor $1/\pi$ of Eq.~\eqref{eq:defs}.
\end{proof}

\paragraph{The vacuum.} For $\rho=|0\rangle\langle0|$ one has $\langle\uu|0\rangle=e^{-|\uu|^2/4}$, hence $Q(\rr)=e^{-|\rr|^2/2}/2\pi$ and
\begin{equation}
\langle\rr+\sss|0\rangle\langle0|\rr-\sss\rangle=e^{-(|\rr+\sss|^2+|\rr-\sss|^2)/4}=e^{-(|\rr|^2+|\sss|^2)/2}=2\pi\sqrt{Q(\rr+\sss)Q(\rr-\sss)} .
\label{eq:vacmatrix}
\end{equation}
\emph{(i) Without the phase.} The right-hand side of Eq.~\eqref{eq:pointwise} is
\begin{equation}
\frac1\pi\int d^2\sss \frac{1}{2\pi}e^{-(|\rr|^2+|\sss|^2)/2}=\frac1\pi e^{-(x^2+p^2)/2},
\end{equation}
since $\int d^2\sss e^{-|\sss|^2/2}=2\pi$. \emph{(ii) With the phase.} Equation~\eqref{eq:parityexplicit} with Eq.~\eqref{eq:vacmatrix} gives
\begin{equation}
W(\rr)=\frac{e^{-|\rr|^2/2}}{\pi}\int\frac{d^2\sss}{2\pi} e^{-|\sss|^2/2} e^{i(x s_p-p s_x)}=\frac{e^{-|\rr|^2/2}}{\pi} e^{-|\rr|^2/2}=\frac1\pi e^{-(x^2+p^2)},
\end{equation}
where we used $\int\frac{ds}{\sqrt{2\pi}}e^{-s^2/2}e^{iks}=e^{-k^2/2}$ once for $s_p$ (with $k=x$) and once for $s_x$ (with $k=-p$). Comparing (i) and (ii): the matrix element~\eqref{eq:vacmatrix} is positive, so the triangle inequality is an equality if and only if the phase $e^{i(x s_p-p s_x)}$ is constant in $\sss$, i.e.\ if and only if $\rr=0$. For $\rr\neq0$ the Fourier transform of the Gaussian in $\sss$ supplies the second factor $e^{-|\rr|^2/2}$, which the bound~\eqref{eq:pointwise} replaces by $1$. Integrating over $\rr$, the bound gives $\int d^2\rr e^{-|\rr|^2/2}/\pi=2$ whereas $\|W\|_1=1$: this is the factor of two between Eq.~\eqref{eq:weak} and Eq.~\eqref{eq:conj}.

\section{The Wigner function as a two-mode wavefunction}
\label{app:twomode}
This Appendix adapts known results from harmonic analysis and signal processing \cite{CORDERO202285,Bayer2020} to quantum information with continuous variables.

We write $|y\rangle$ for position eigenkets, $\langle y|y'\rangle=\delta(y-y')$. For a single-mode pure state $|\psi\rangle$ we define a state $|\bar\psi\rangle$,
\begin{equation}
|\bar\psi\rangle=\int dy \langle y|\bar\psi\rangle |y\rangle=\int dy \overline{\langle y|\psi\rangle} |y\rangle=\int dy \langle\psi|y\rangle |y\rangle.
\end{equation}
That implies that the position wavefunction of state $|\bar\psi\rangle$ is $\bar{\psi}(y)$, i.e., the complex conjugate of the the position wavefunction of $|\psi\rangle$, or equivalently $\langle\psi|y\rangle=\langle y|\bar\psi\rangle$.

The balanced beam splitter $\hat B$ and the Fourier transform $\hat F$ act on position eigenkets as
\begin{equation}
\hat B |x,y\rangle=\Big|\frac{x+y}{\sqrt2},\frac{x-y}{\sqrt2}\Big\rangle,
\qquad
\hat F|y\rangle=\frac{1}{\sqrt{2\pi}}\int dx e^{iyx} |x\rangle.
\label{eq:BS}
\end{equation}

We define
\begin{equation}
|\Xi\rangle=\hat V\left(|\bar\psi\rangle\otimes|\psi\rangle\right),\qquad \hat V=(\hat 1\otimes\hat F) \hat B ,
\qquad \Xi(x,y)=\langle x,y|\Xi\rangle,
\label{eq:Xidef}
\end{equation}
where $|\Xi\rangle$ is normalized and $\hat V$ is passive.

The Wigner function for a pure state $|\psi\rangle$ is,
\begin{equation}
W(x,p)=\frac1\pi\int dy e^{2ipy} \langle\psi|x+y\rangle \langle x-y|\psi\rangle
=\frac1\pi\int dy e^{2ipy} \langle x+y, x-y|\big(|\bar\psi\rangle\otimes|\psi\rangle\big),
\label{eq:Wposition}
\end{equation}
where we used $\langle\psi|x+y\rangle=\langle x+y|\bar\psi\rangle$. The single-mode Wigner function is thus an integral of a two-mode position amplitude of the product state $|\bar\psi\rangle\otimes|\psi\rangle$. We can proceed further to reveal more structure as follows:\\
\noindent
\emph{Balanced beam splitter.} By Eq.~\eqref{eq:BS} and the fact that $\hat{B}^\dagger=\hat{B}$ we get that $\langle\sqrt2x,\sqrt2y|\hat B=\langle x+y, x-y|$ and by defining $y'=\sqrt2y$ we get,
\begin{equation}
W(x,p)=\frac{1}{\sqrt2\pi}\int dy' e^{i\sqrt2p y'}\langle\sqrt2x,y'|\hat B\left(|\bar\psi\rangle\otimes|\psi\rangle\right).
\end{equation}
\emph{Fourier transform.} By Eq.~\eqref{eq:BS} $\int dy' e^{ip'y'}\langle y'|=\sqrt{2\pi} \langle p'|\hat F$ and by defining $p'=\sqrt2p$,
\begin{equation}
W(x,p)=\frac{\sqrt{2\pi}}{\sqrt2 \pi} \langle\sqrt2x,\sqrt2p| (\hat1\otimes\hat F) \hat B\left(|\bar\psi\rangle\otimes|\psi\rangle\right)
=\frac{1}{\sqrt\pi} \langle\sqrt2x,\sqrt2p|\hat V\left(|\bar\psi\rangle\otimes|\psi\rangle\right).
\end{equation}
We note that both $p$ and $p'$ carry the meaning of position eigenvectors.

Hence the Wigner function of $|\psi\rangle$ is, up to a constant and a rescaling of its arguments, the position wavefunction of $|\bar\psi\rangle\otimes|\psi\rangle$ transformed by $\hat V$,
\begin{equation}
W(x,p)=\frac{1}{\sqrt\pi} \Xi\big(\sqrt2 x,\sqrt2 p\big),
\qquad
\|W\|_1=\frac{1}{2\sqrt\pi} \|\Xi\|_{L^1(\mathbb R^2)} ,
\label{eq:WXi}
\end{equation}
where the second relation follows from the change of variables $(X,P)=\sqrt2(x,p)$.

The factors $\sqrt2$ in Eq.~\eqref{eq:WXi} only rescale the point at which $\Xi$ is evaluated: they are a change of integration variables, not a squeezing of the state, which is obtained from $|\bar\psi\rangle\otimes|\psi\rangle$ by the passive unitary $\hat V$ alone. Note also that Eq.~\eqref{eq:WXi} involves the amplitude $\Xi(X,P)$ itself, including its phase, which is why $|\Xi\rangle$ must be defined as a state vector and not only through $|\Xi\rangle\langle\Xi|$. As a consistency check, $\Xi$ is real because $W$ is, and $\int dx dp W^2=\frac{1}{2\pi}\langle\Xi|\Xi\rangle=\frac{1}{2\pi}$, as it should for a pure state.

\section{Passive operations preserve the coherent-amplitude norms}
\label{app:passive}

For an $n$-mode pure state $\Phi$ we write $|\boldsymbol\alpha\rangle=|\alpha_1\rangle\otimes\cdots\otimes|\alpha_n\rangle$ and define the \emph{coherent amplitude} $f_\Phi(\boldsymbol\alpha)=\langle\boldsymbol\alpha|\Phi\rangle$ and its $L^1$ norm
\begin{equation}
N(\Phi)=\int\frac{d^{2n}\alpha}{\pi^n} |f_\Phi(\boldsymbol\alpha)|=\int\prod_{i=1}^n\frac{d^2\rr_i}{2\pi} \big|\langle\rr_1,\dots,\rr_n|\Phi\rangle\big| ,
\label{eq:Ndef}
\end{equation}
where $d^2\alpha=dx dp/2$. For a single mode, $|f_\psi(\alpha)|^2=2\pi Q(\rr)$, so
\begin{equation}
N(\psi)=\frac{1}{\sqrt{2\pi}}\int d^2\rr \sqrt{Q(\rr)} ,\qquad N(|0\rangle)=2 .
\label{eq:N1}
\end{equation}

\begin{lemma}[Passive invariance]
\label{lem:passive}
Let $\hat V$ be a passive unitary, i.e.\ one that maps annihilation operators to linear combinations of annihilation operators,
\begin{equation}
\hat V^\dagger\hat a_i\hat V=\sum_{j=1}^nU_{ij} \hat a_j ,\qquad U\ \text{an}\ n\times n\ \text{unitary matrix}.
\label{eq:passive}
\end{equation}
Then (a) $\hat V|0\rangle=e^{i\theta}|0\rangle$, (b) $\hat V^\dagger|\boldsymbol\alpha\rangle=e^{-i\theta}|U^\dagger\boldsymbol\alpha\rangle$ for every $\boldsymbol\alpha\in\mathbb C^n$, (c) $f_{\hat V\Phi}(\boldsymbol\alpha)=e^{i\theta}f_\Phi(U^\dagger\boldsymbol\alpha)$ and (d) $\int d^{2n}\alpha |f_{\hat V\Phi}|^q=\int d^{2n}\alpha |f_\Phi|^q$ for every $q>0$. In particular $N(\hat V\Phi)=N(\Phi)$.
\end{lemma}

\begin{proof}
(a) For every $i$, $\hat a_i\hat V|0\rangle=\hat V(\hat V^\dagger\hat a_i\hat V)|0\rangle=\hat V\sum_jU_{ij}\hat a_j|0\rangle=0$. The vacuum is the unique state (up to a phase) annihilated by all $\hat a_i$, so $\hat V|0\rangle=e^{i\theta}|0\rangle$, and then also $\hat V^\dagger|0\rangle=e^{-i\theta}|0\rangle$.

(b) The $n$-mode displacement operator is $\hat D(\boldsymbol\alpha)=\exp\big[\sum_i(\alpha_i\hat a_i^\dagger-\bar\alpha_i\hat a_i)\big]$. Conjugation by a unitary acts inside the exponential, and Eq.~\eqref{eq:passive} with its adjoint $\hat V^\dagger\hat a_i^\dagger\hat V=\sum_j\bar U_{ij}\hat a_j^\dagger$ give
\begin{equation}
\hat V^\dagger\hat D(\boldsymbol\alpha)\hat V=\exp\Big[\sum_j\Big(\sum_i\bar U_{ij}\alpha_i\Big)\hat a_j^\dagger-\mathrm{h.c.}\Big]=\hat D(U^\dagger\boldsymbol\alpha),
\end{equation}
with no additional phase, because $(U^\dagger\boldsymbol\alpha)_j=\sum_i\bar U_{ij}\alpha_i$. Therefore $\hat V^\dagger|\boldsymbol\alpha\rangle=\hat V^\dagger\hat D(\boldsymbol\alpha)\hat V \hat V^\dagger|0\rangle=e^{-i\theta}\hat D(U^\dagger\boldsymbol\alpha)|0\rangle$.

(c) $f_{\hat V\Phi}(\boldsymbol\alpha)=\langle\boldsymbol\alpha|\hat V|\Phi\rangle=\overline{\langle\Phi|\hat V^\dagger|\boldsymbol\alpha\rangle}=e^{i\theta}\langle U^\dagger\boldsymbol\alpha|\Phi\rangle$.

(d) Substitute $\boldsymbol\beta=U^\dagger\boldsymbol\alpha$. A unitary matrix on $\mathbb C^n$ preserves the Euclidean norm $\sum_i|\alpha_i|^2$ of $\mathbb R^{2n}$, so it is a real orthogonal map of $\mathbb R^{2n}$ and has unit Jacobian: $d^{2n}\beta=d^{2n}\alpha$.
\end{proof}

A passive operation therefore only relabels the argument of the coherent amplitude by a rotation of phase space. That is, all its $L^q$ norms, and hence all R\'enyi--Wehrl entropies of the Husimi function, are unchanged. This fails for active (squeezing) operations: they mix $\hat a$ with $\hat a^\dagger$, do not preserve the vacuum, and map coherent states to squeezed states, so that $\langle\boldsymbol\alpha|\hat S|\Phi\rangle$ is an overlap with a squeezed state and not $f_\Phi$ at a displaced point.

\begin{lemma}[Complex conjugation]
\label{lem:conj}
$|f_{\bar\psi}(\alpha)|=|f_\psi(\bar\alpha)|$, and consequently $N(\bar\psi)=N(\psi)$.
\end{lemma}

\begin{proof}
From Eq.~\eqref{eq:cohwf}, $\overline{\langle y|(x,p)\rangle}=\langle y|(x,-p)\rangle$. Hence
\begin{equation}
f_{\bar\psi}(x,p)=\int dy \overline{\langle y|(x,p)\rangle} \overline{\psi(y)}=\overline{\int dy \overline{\langle y|(x,-p)\rangle} \psi(y)}=\overline{f_\psi(x,-p)} .
\end{equation}
The reflection $p\to-p$ has unit Jacobian.
\end{proof}

\begin{corollary}[Reformulation]
\label{cor:reform}
For $\Xi$ as in Eq.~\eqref{eq:Xidef},
\begin{equation}
N(\Xi)=N(\bar\psi\otimes\psi)=N(\psi)^2=\frac{1}{2\pi}\Big(\int d^2\rr \sqrt{Q(\rr)}\Big)^2 .
\label{eq:NXi}
\end{equation}
Consequently, for pure states, the Conjecture~\eqref{eq:conj} and Proposition~\ref{prop:weak} are respectively equivalent to
\begin{equation}
\|\Xi\|_{L^1(\mathbb R^2)}\le\frac{\sqrt\pi}{2} N(\Xi)
\qquad\text{and}\qquad
\|\Xi\|_{L^1(\mathbb R^2)}\le\sqrt\pi N(\Xi).
\label{eq:reform}
\end{equation}
\end{corollary}

\begin{proof}
The first equality in Eq.~\eqref{eq:NXi} is Lemma~\ref{lem:passive} applied to the passive unitary $(\hat1\otimes\hat F)\hat B$. The coherent amplitude of a product state factorizes, $f_{\bar\psi\otimes\psi}(\alpha_1,\alpha_2)=f_{\bar\psi}(\alpha_1)f_\psi(\alpha_2)$, so $N(\bar\psi\otimes\psi)=N(\bar\psi)N(\psi)=N(\psi)^2$ by Lemma~\ref{lem:conj}, and the last equality is Eq.~\eqref{eq:N1}. For Eq.~\eqref{eq:reform}, insert $\|W\|_1=\|\Xi\|_{L^1}/2\sqrt\pi$ from Eq.~\eqref{eq:WXi} and $(\int\!\sqrt Q)^2=2\pi N(\Xi)$ into Eqs.~\eqref{eq:conj} and~\eqref{eq:weak}: $\frac{1}{2\sqrt\pi}\|\Xi\|_{L^1}\le\frac{2\pi}{8\pi}N(\Xi)$ and $\frac{1}{2\sqrt\pi}\|\Xi\|_{L^1}\le\frac{2\pi}{4\pi}N(\Xi)$.
\end{proof}

The two constants in Eq.~\eqref{eq:reform} are products of single-mode constants. For the single-mode vacuum, $\psi_0(y)=\pi^{-1/4}e^{-y^2/2}$, one has $\|\psi_0\|_{L^1(\mathbb R)}=\sqrt2 \pi^{1/4}$ and $N(\psi_0)=2$, so the ratio is $\pi^{1/4}/\sqrt2$ while its square is $\sqrt\pi/2$ (the conjectured constant). The two-mode vacuum thus saturates the first inequality in Eq.~\eqref{eq:reform}, consistently with $\|W\|_1=1$ for the vacuum.

\section{The constant of Proposition~\ref{prop:weak} is optimal for arbitrary wavefunctions}
\label{app:sharp}

\begin{lemma}[General bound]
\label{lem:general}
For every $n$-mode wavefunction $\Phi$ with $N(\Phi)<\infty$,
\begin{equation}
\|\Phi\|_{L^1(\mathbb R^n)}\le\pi^{n/4} N(\Phi).
\label{eq:general}
\end{equation}
\end{lemma}

\begin{proof}
Consider first $n=1$. By Eq.~\eqref{eq:cohwf},
\begin{equation}
f_\phi(x,p)=\langle(x,p)|\phi\rangle=\pi^{-1/4}\int dy' e^{-(y'-x)^2/2-ipy'+ipx/2} \phi(y').
\end{equation}
Set $x=y$, multiply by $e^{ipy/2}$ and integrate over $p$:
\begin{equation}
\int\frac{dp}{2\pi} e^{ipy/2}f_\phi(y,p)=\pi^{-1/4}\int dy' e^{-(y'-y)^2/2}\phi(y')\int\frac{dp}{2\pi}e^{ip(y-y')}=\pi^{-1/4}\phi(y).
\end{equation}
Hence $\phi(y)=\pi^{1/4}\int\frac{dp}{2\pi}e^{ipy/2}f_\phi(y,p)$, so $|\phi(y)|\le\pi^{1/4}\int\frac{dp}{2\pi}|f_\phi(y,p)|$, and integrating over $y$ gives $\|\phi\|_{L^1}\le\pi^{1/4}N(\phi)$. For $n$ modes the same reconstruction formula holds in each variable separately, which gives Eq.~\eqref{eq:general}.
\end{proof}

For $n=2$, Eq.~\eqref{eq:general} is the second inequality in Eq.~\eqref{eq:reform}. Lemma~\ref{lem:general} together with Corollary~\ref{cor:reform} is therefore a second proof of Proposition~\ref{prop:weak} for pure states, one that uses no property of $\Xi$ other than its being a two-mode wavefunction.

\begin{lemma}[Squeezed vacuum]
\label{lem:squeezed}
Let $\psi_s(y)=(\pi s^2)^{-1/4}e^{-y^2/2s^2}$, $s>0$, be the squeezed vacuum of position width $s$. Then
\begin{equation}
\|\psi_s\|_{L^1(\mathbb R)}=\sqrt2 \pi^{1/4}\sqrt s,\qquad
N(\psi_s)=\sqrt{\frac{2(1+s^2)}{s}},\qquad
\frac{\|\psi_s\|_{L^1}}{N(\psi_s)}=\pi^{1/4}\frac{s}{\sqrt{1+s^2}} .
\label{eq:sqratio}
\end{equation}
\end{lemma}

\begin{proof}
The first equality is $\int dy e^{-y^2/2s^2}=\sqrt{2\pi} s$. For the second, by Eq.~\eqref{eq:cohwf},
\begin{equation}
f_{\psi_s}(x,p)=\frac{e^{ipx/2}}{\pi^{1/4}(\pi s^2)^{1/4}}\int dy \exp\!\Big[-a y^2+(x-ip) y-\frac{x^2}{2}\Big],\qquad a=\frac{1+s^2}{2s^2}.
\end{equation}
The Gaussian integral equals $\sqrt{\pi/a} \exp[(x-ip)^2/4a-x^2/2]$, with modulus $\sqrt{\pi/a} \exp[(x^2-p^2)/4a-x^2/2]$. Using $1/4a=s^2/2(1+s^2)$,
\begin{equation}
|f_{\psi_s}(x,p)|=\sqrt{\frac{2s}{1+s^2}} \exp\!\Big[-\frac{x^2}{2(1+s^2)}-\frac{s^2p^2}{2(1+s^2)}\Big],
\end{equation}
and $N(\psi_s)=\frac{1}{2\pi}\sqrt{\frac{2s}{1+s^2}}\cdot\sqrt{2\pi(1+s^2)}\cdot\frac{\sqrt{2\pi(1+s^2)}}{s}=\sqrt{2(1+s^2)/s}$. The ratio follows.
\end{proof}

\begin{corollary}[Optimality]
\label{cor:optimal}
(a) The ratio in Eq.~\eqref{eq:sqratio} exceeds the vacuum value $\pi^{1/4}/\sqrt2$ if and only if $s>1$, is strictly increasing in $s$, and tends to $\pi^{1/4}$ as $s\to\infty$. (b) The constant in Eq.~\eqref{eq:general} cannot be improved: $\sup_\Phi\|\Phi\|_{L^1(\mathbb R^n)}/N(\Phi)=\pi^{n/4}$, the supremum being approached by $\psi_s^{\otimes n}$ as $s\to\infty$ and not attained by it for any finite $s$. (c) In particular, the first inequality in Eq.~\eqref{eq:reform} is false for arbitrary two-mode wavefunctions: $\psi_s\otimes\psi_s$ violates it for every $s>1$.
\end{corollary}

\begin{proof}
(a) $s^2/(1+s^2)$ is strictly increasing from $0$ to $1$ and equals $1/2$ at $s=1$. (b) Both $\|\cdot\|_{L^1}$ and $N$ are multiplicative on product states, so $\psi_s^{\otimes n}$ has ratio $\pi^{n/4}\big(s^2/(1+s^2)\big)^{n/2}\to\pi^{n/4}$, which is the upper bound of Lemma~\ref{lem:general}. (c) For $n=2$ the ratio is $\sqrt\pi s^2/(1+s^2)$, which exceeds $\sqrt\pi/2$ exactly when $s>1$.
\end{proof}

By Corollary~\ref{cor:reform}, the constant $\sqrt\pi$ in Eq.~\eqref{eq:general} for $n=2$ corresponds to the constant $1/4\pi$ of Proposition~\ref{prop:weak}, and the constant $\sqrt\pi/2$ to the conjectured $1/8\pi$. Corollary~\ref{cor:optimal} thus shows that, over the set of all two-mode wavefunctions, the best possible constant is the one of Proposition~\ref{prop:weak}. Any proof of the Conjecture must therefore use that $\Xi$ is of the special form~\eqref{eq:Xidef}.

Finally, the wavefunctions $\psi_s\otimes\psi_s$ that approach the constant of Proposition~\ref{prop:weak} are not of the form~\eqref{eq:Xidef} unless $s=1$. Indeed, if $\Xi=\psi_s\otimes\psi_s$ arose from some pure state through Eq.~\eqref{eq:Xidef}, then Eq.~\eqref{eq:WXi} would give
\begin{equation}
W(x,p)=\frac{1}{\sqrt\pi} \psi_s(\sqrt2x) \psi_s(\sqrt2p)=\frac{1}{\pi s} e^{-(x^2+p^2)/s^2},
\qquad
\int dx dp W(x,p)=s ,
\end{equation}
which contradicts the normalization $\int W=1$ for $s\neq1$. Equivalently, $\int dX dP \Xi(X,P)=2\sqrt\pi$ for every $\Xi$ of the form~\eqref{eq:Xidef}, whereas $\int\psi_s\otimes\psi_s=2\sqrt\pi s$. For $s>1$ this function is a Gaussian that is broad in both $x$ and $p$ and is correctly normalized, $e^{-(x^2+p^2)/s^2}/\pi s^2$ is the Wigner function of a thermal state of purity $1/s^2<1$, not of a pure state. The normalization and purity constraints that a pure-state Wigner function must satisfy are exactly what excludes the wavefunctions responsible for the constant $1/4\pi$.

\twocolumngrid

\bibliography{refs}

\end{document}